\documentclass[12pt,a4paper]{article}%
\usepackage[utf8]{inputenc}
\usepackage{amsmath}
\usepackage{amsfonts}
\usepackage{amssymb}
\usepackage{graphicx}
\usepackage[margin=1.25in]{geometry}
\usepackage{makecell}
\usepackage{chngcntr}

\usepackage{mathrsfs}
\usepackage{}
\usepackage{multirow}
\usepackage{subfigure}
\usepackage{float}
\usepackage{authblk}

\usepackage[
  backend=bibtex,%biber,
  natbib=true,
  style=numeric,
  sorting=none,
  maxbibnames=99,
  firstinits=true]{biblatex}%backend=bibtex,   
\usepackage[usenames,dvipsnames]{color}
\usepackage{url}
\def\mm {\mu^+\mu^-}

\newcommand{\mz}{M_{Z}}
\newcommand{\mh}{M_{h}}
\newcommand{\Mmu}{M_{\mu^{+}\mu^{-}}}
\newcommand{\ptmu}{P_{\mathrm{T}}^{\mu^{+}\mu^{-}}}
\newcommand{\Mrec}{M_{\mathrm{rec}}}

\newcommand{\cosmis}{\text{cos}\theta_{\text{mis}}}

\newcommand{\pb}{\,\text{pb}}
\newcommand{\fb}{\,\text{fb}}
\newcommand{\gev}{\,\text{GeV}}

\newenvironment{Abstract}{\begin{quotation} \begin{center}
                       ABSTRACT
     \end{center}\bigskip  }{\end{quotation}}

\def\Acknowledgements{\bigskip  \bigskip \begin{center} 
             \bf ACKNOWLEDGEMENTS \end{center}}

\vfill
\title{Search for Light Scalars Produced in Association with Muon Pairs for $\sqrt{s}$ = 250 GeV at the ILC}
\vfill

\author[1,2]{YAN WANG\thanks{yan.wang@desy.de}}
\author[1]{JENNY LIST}
\author[1]{MIKAEL BERGGREN}
\affil[1]{DESY, Notkestra\ss e 85, 22607 Hamburg, GERMANY}
\affil[2]{IHEP, Yuquan Road 19B, 100049 Beijing, CHINA}
\date{\today}

\begin{document}
 
%\date{\today}

\maketitle

\vfill
\begin{Abstract}

In many models with extended Higgs sectors, $e.g.$ 2HDM, NMSSM, there exists a light scalar $h$, lighter than the  Standard Model (SM) like Higgs, and the coupling of $hZZ$ can be very small, as expected from the likeness of the 125 GeV Higgs boson measured at the LHC to the SM Higgs boson. Such a light scalar with suppressed couplings to the $Z$ boson would have escaped detection at LEP due to its limited luminosity. With a factor of 1000 higher luminosity and polarized beams, the International Linear Collider (ILC) is expected to have substantial discovery potential for such states. Furthermore, searches for additional scalars  at LEP and LHC are usually dependent on the model details, such as decay channels. Thus, it is necessary to have a more general analysis with model-independent assumptions.

    In this work, we perform a search for a light higgs boson produced in association with $Z$ boson at the ILC with a center-of-mass energy of 250 GeV, using the full Geant4-based simulation of the ILD detector concept. In order to be as model-independent as possible, the analysis is performed using the recoil technique, in particular with the $Z$ boson decaying into a pair of muons. As a preliminary result, the ILC's exclusion limits will be shown for different higgs masses between 30 and 115 GeV.\footnote[2]{Talk presented at the International Workshop on Future Linear Colliders (LCWS2017), Strasbourg, France, 23-27 October 2017. C17-10-23.2.}
\end{Abstract}
\vfill

\section{Introduction}

Many new physics models predict one or more extra scalars. For example, in Two Higgs Doublet Model (2HDM), the additional neutral particles are two scalars and a pseudoscalar \cite{PhysRevD.8.1226,Branco:2011iw}. In the Next-to-Minimal Supersymmetric Standard Model (NMSSM), it has three scalars and two pseudoscalars \cite{FAYET1975104,FAYET1977,FAYET1977489,PhysRevD.39.844,DREES1989}. In these models, a scalar lighter than $125\gev$ is well motivated.
However, the $125\gev$ Higgs boson measured at the LHC is rather Standard Model (SM) like \cite{Khachatryan:2016vau, deFlorian:2016spz}\footnote{In this paper, "higgs" is used for a scalar lighter than $125\gev$, while "Higgs" stands for the SM/$125\gev$ Higgs boson.}. 
Thus, if such a new light higgs exists, its coupling to the $Z$ boson will be suppressed \cite{Aggleton:2016tdd}.
The LEP searches for the SM Higgs boson can be used to constrain these additional scalars, but in most cases only when their properties, especially decay profiles, are similar to those of the SM Higgs boson \cite{Barate:2003sz,Sguazzoni:2004xc, 1742-6596-110-4-042030}. Furthermore, LEP/LHC constraints on the extra scalars rely on the model details, $e.g.$ CP properties, mass hierarchy, couplings, $etc.$ \cite{Cacciapaglia:2016tlr}.  
Therefore, it is necessary to have a more general analysis with as few assumptions as possible. The OPAL collaboration has searched for light scalars in a model-independent way at LEP, but the results are limited by luminosity \cite{Abbiendi:2002qp}. 

The International Linear Collider (ILC) is an electron-positron linear collider, with a center-of-mass energy of $250\gev$ at its first stage.
  In Table \ref{t_compareLEP}, there is a brief comparison between LEP and ILC when searching for scalars \cite{Barate:2003sz, Abbiendi:2002qp,Behnke:2013lya}.  Comparing with LEP, ILC has 1000 times higher luminosity, which makes the recoil mass technique more accurate, and it can provide more observables with polarized beams, such as angular correlations, which will help to distinguish unique signals of the scalar production \cite{Asner:2013psa, Yan:2016xyx}.
Thus, ILC will be sensitive to light scalars with a very weak interaction of $Z$ boson using model-independent analysis.

\begin{table}[h]\label{t_compareLEP}
\begin{center}
 \begin{tabular}{|c|c|c|} \hline
 ~~ & LEP & ILC  \\
 \hline
 $\sqrt{s}$  & \thead{LEP1: $91.2\gev$ \\ LEP2: $189~\text{to}~209\gev$} & $250\gev$ \\
 \hline
 Beam polarization & No & Yes \\
  \hline
Integrated luminosity  & \thead{traditional method: \\2461 $\pb^{-1}$ for $\sqrt{s} \geq 189\gev$ \\ 536 $\pb^{-1}$ for $\sqrt{s} \geq 206\gev$ \vspace{0.2cm} \\   \hline \\ recoil method: \\$115.4 \pb^{-1} \text{at LEP1}$ \\ $662.4 \pb^{-1} \text{at LEP2}$ } & $2000\fb^{-1}$ \\
 \hline

 Search channel & \thead{traditional method:  \\ $2b2q$,$2b2\nu$,$2b2l$, $\tau\tau qq$ \vspace{0.2cm} \\   \hline \\ recoil method: \\model independent}  & model independent \\
 \hline
 Experiment ingredient & \thead{traditional method: \\$b$-tagging \vspace{0.2cm} \\  \hline \\ recoil method: \\ recoil mass}
 &   \thead{ recoil mass \\ angle correlation \\ momentum resolution}
\\
 \hline
 \end{tabular}
 \caption{Comparison of LEP and ILC characteristics when searching scalars. The traditional method for discriminating signal and background at LEP is identifying the decay modes of scalars \cite{Barate:2003sz}. The recoil method at LEP refers to the analysis with recoil technique by the OPAL collaboration at LEP \cite{Abbiendi:2002qp}.}
  \end{center}
\end{table}

This paper is structured as follows: Section \ref{s_simulation} introduces the signal and background processes as well as  the ILD detector concept and the detector simulation tools; Section \ref{s_analysis} presents the methods of data selection; Section \ref{s_result} shows the exclusion limits for $hZZ$ coupling; Section \ref{s_conclusion} concludes the paper.  

\section{Event Generation and Detector Simulation}\label{s_simulation}

\subsection{Signal and background processes}

We consider here the production of a light scalar $h$ in association with a $Z$ boson where the $Z$ boson decays to a muon pair. 
%The Feynman diagram is shown in Figure \ref{f_feynman}, which is similar to the SM higgsstrahlung process.  
The $\mu\mu h$ signal Monte Carlo (MC) samples have been generated using the Whizard 1.95 event generator \cite{Kilian2011}, at a center-of-mass energy of 250 GeV, for 100\% left-handed and right-handed beam polarization configurations. 
Five benchmark points are chosen for the signal MC samples, $\mh=30, 50, 70, 90, 115 \gev$. The decay branching ratios are assumed to be the same as a $125\gev$ SM Higgs boson, but no use would be made of this fact.

\begin{figure}[h]
  \begin{center}
     \includegraphics[height=6cm]{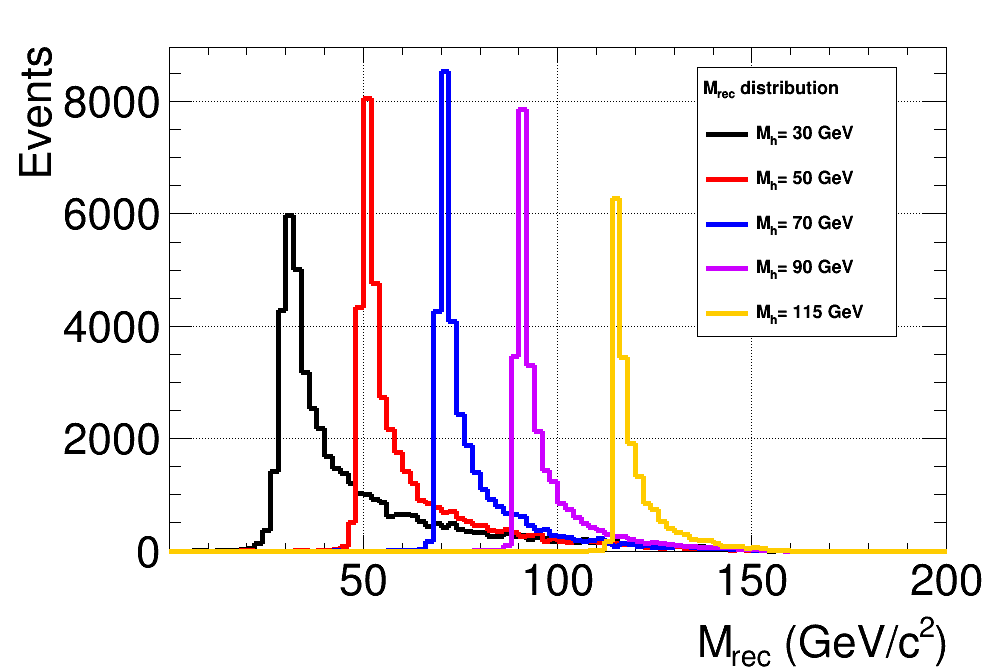}	
  \end{center}
     \caption{The recoil mass distribution of the signal for different scalar masses, when integrated luminosity is 2000 $\fb^{-1}$, $P(e^{-},e^{+})=(-80\%,+30\%)$, and $\sqrt{s}=250\gev$. }\label{f_recoil}
\end{figure}

The scalar mass can be measured by the recoil technique due to the high luminosity at the ILC. 
  Figure \ref{f_recoil} shows the recoil mass distributions ($\Mrec$) for different scalar masses, where the peak of recoil mass distributions are very sharp for each mass, which can be used to distinguish the signal and background. 

As backgrounds, we use samples which have been generated in the context of the Detailed Baseline Design document (ILC TDR, Vol. 4 \cite{,Behnke:2013lya, dbdsample}). They are generated at tree level and grouped by lepton numbers in the final state as follows:
\begin{itemize}
\item 2-fermion leptonic ($2f_l$), main channel: $e^{+}e^{-} \to Z/\gamma^{*} \to l^{+}l^{-}/\nu\nu$. 

\item 4-fermion leptonic ($4f_l$), main channel: $e^{+}e^{-} \to ZZ/WW \to 4l$.

\item 4-fermion semi-leptonic ($4f_{sl}$), main channel: $e^{+}e^{-} \to ZZ/WW \to 2l2q/2\nu2q$.

\item 4(2)-fermion hadronic ($4f_{h}~/~2f_{h}$), main channel: $e^{+}e^{-} \to ZZ/WW \to 4q$ or $e^{+}e^{-} \to Z \to 2q$. These events are almost fully rejected when identifying isolated muon pairs. 
\end{itemize}

Bremsstrahlung and initial state radiation (ISR) are explicitly taken into account for all events. 
Pythia 6.4 is used for the parton shower and hadronization  \cite{Sjostrand:2006za}. The samples are reweighted to the polarization of $P(e^{-},e^{+})$ = $(-80\%,\,+30\%)$, and an integrated luminosity of 2000 $\fb^{-1}$.

\subsection{Detector simulation}
The generated events have been simulated with the full Geant4-based \cite{Agostinelli:2002hh} simulation of the ILD detector concept \cite{Behnke:2013lya},  more specifically the ILD$\_$o1$\_$v05 detector modle in mokka \cite{MoradeFreitas:2002kj}.  The simulated events have been reconstructed with the standard tools in ILCSoft v01-16 \cite{ILCSOFT}.
The ILD detector is designed for optimal particle-flow performance \cite{Abe:2010aa}. It has a vertex detector consisting of three double-sided layers of silicon pixel sensors, a hybrid tracking system, which is realized with a time projection chamber and a combination of silicon tracking,  and a calorimeter system. These systems are surrounded by a solenoid producing a 3.5 T magnetic field, and an iron flux return yoke.
Event reconstruction has been performed  using the PandoraPFA algorithm \cite{THOMSON200925} to reconstruct individual particles  within the Marlin framework \cite{GAEDE2006177}.	The beam crossing angle of 14 mrad has been also taken into account \cite{Behnke:2013lya}.

\section{Analysis}\label{s_analysis}
 
In this section, we describe the analysis strategy following the cross section measurement of the SM Higgs boson at the ILC \cite{Yan:2016xyx}. 
The signal is selected by firstly identifying a pair of isolated and oppositely charged muons. Then final state radiation (FSR) and bremsstrahlung photons are recovered. Finally, background events are rejected with several kinematic cuts.
\subsection{Selection of the best muon pair}
Isolated muons are identified with the following criteria.
\begin{itemize}
\item Muons are required to have sufficient track momentum:  $p_{\text{track}} > 5\gev$, where $p_{\text{track}}$ is the measured track momentum.
\item Muon ionization is minimal when passing the ECAL and HCAL: \\$E_{\text{CAL,tot}}/p_{\text{track}} < 0.3$, $E_{\text{yoke}} > 1.2\gev$, where $E_{\text{CAL,tot}}$ and $E_{\text{yoke}}$ is the energy deposit in ECAL plus HCAL and inside the muon detector.
\item  Suppression of the muons from $\tau$ decay or $b/c$ quark jets by requiring the uncertainties of $d_0$ and $z_0$, $|d_{0}/\delta d_{0}| < 5$, $|z_{0}/\delta z_{0}| < 5$, where $d_0$ and $z_0$ are the impact parameters in the transverse and longitudinal direction and $\delta d_{0}$ and $\delta z_{0}$ are their uncertainties.
\end{itemize}
Then a multi variate method is used for further identifying isolated muons \cite{GAEDE2006177,leptag}. 

For the signal, at least one pair of isolated oppositely charged muons is selected. 
However, there may be more muons in the events, for example, the muons produced in the higgs decay process $h \to WW^{*}$ and $h \to ZZ^{*}$. In order to find the correct muon pair, which is produced from the $Z$ boson decay in the associated production process, the invariant mass $\Mmu$ of the correct muon pair should be close to the $Z$ boson mass $\mz$=91.2 GeV. Meanwhile, with the recoil technique, the recoil mass of the correct muon pair system can be calculated from Eq. (\ref{eq_recoil}). It should be close to higgs boson mass $\mh=30,~50,~70,~90,~\text{or}~115\gev$, respectively, for each higgs mass benchmark point in this study: 
\begin{equation}\label{eq_recoil}
\Mrec^2=(\sqrt{s}-E_{\mm})^2-|\vec{p}_{\mm}|^2.
\end{equation}

Thus the best muon pair candidate is selected with the following criteria: First, the basic criteria $|\Mmu -M_{Z}| <$ 40 GeV is applied. Then the muon pair should minimize the following $\chi^{2}$ function:
\begin{equation}
\chi^2 (\Mmu,\Mrec) = \frac{(\Mmu - \mz)^{2}}{\sigma_{\Mmu}^{2}} + \frac{(\Mrec - \mh)^{2}}{\sigma_{\Mrec}^{2}}.
\end{equation}
where $\sigma_{\Mmu}$ and $\sigma_{\Mrec}$ are determined by a Gaussian
fit to the generator-level distributions of $\Mmu$ and $\Mrec$. 

After identifying the best muon pair, the bremsstrahlung and FSR photons from the muons are identified and added to the muons. A photon is identified as a FSR or bremsstrahlung photon when its cosine of the polar angle with respect to the isolated muons exceeds 0.99. Then the four momentum of the photon is combined with that muon.
\subsection{Background rejection}
Background events are rejected by considering kinematic variables as described below, while the specific cut values are adjusted for each higgs mass. 
\begin{itemize}

\item Since $\Mmu$ should be close to the $Z$ boson mass, a criterion is imposed as $\Mmu\in [73, 120]\gev$ for each higgs mass. Figure \ref{f_zmass} compares the $\Mmu$ distribution of signal and background processes when $\mh=50\gev$.
\begin{figure}[h]
  \begin{center}
     \includegraphics[height=6cm]{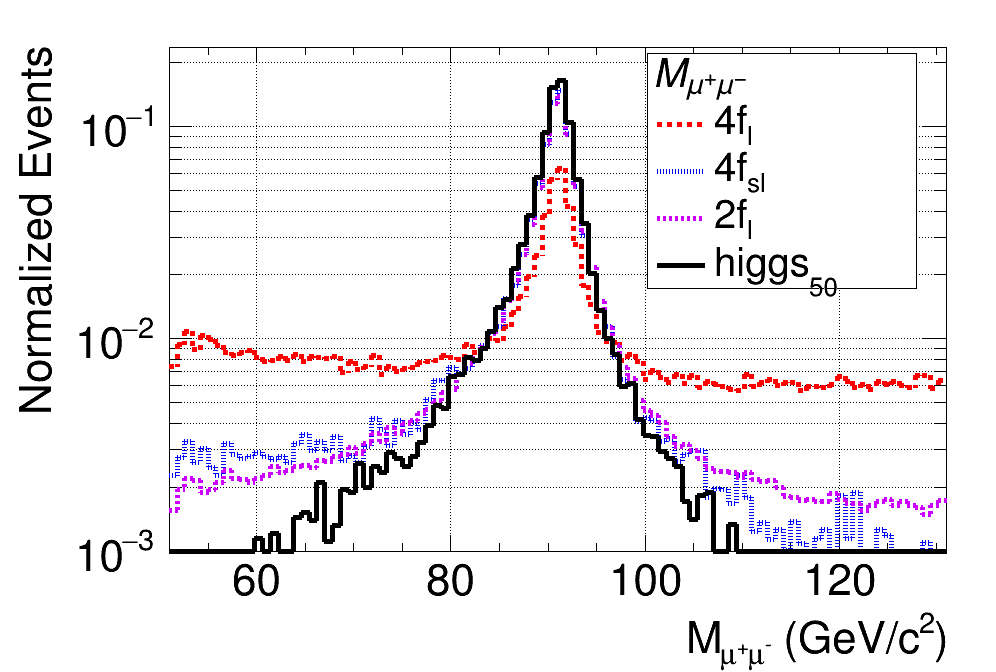}	\\	
  \end{center}
   \caption{Invariant mass distribution of muon pair system for signal $\mh=50\gev$ and backgrounds.}\label{f_zmass}
 \end{figure}

\item  The transverse momentum of the muon pair system $\ptmu$ in the $2f_l$  channel tends to have very small values, in contrast to the signal, which should have a peak at larger values which is determined by kinematics. This motivates $\ptmu > 10\gev$.  In addition, an upper limit is also needed to reduce other backgrounds with large transverse momentum, mainly the events from $2f_l$ processes. In order to maximize the sensitivity, the upper limit of  $\ptmu$ for each higgs mass benchmark point is chosen as the maximum $\ptmu$  value in the region $|\mh^{i}-\mh^{\mathrm{benchmark}}|<10\gev$ . The maximum $\ptmu$ is calculated as a function of the higgs mass, as shown in Figure \ref{f_zpt} (a), and the $\ptmu$ upper limit cuts are shown in Table \ref{t_cut_zpt}. When $\mh<50\gev$, the maximum $\ptmu$ cut is not necessary.
As an example, the Figure \ref{f_zpt} (b) compares the $\ptmu$ distribution of the signal and major background processes when $\mh=~50\gev$.

\begin{figure}[h] 

  \begin{minipage}[t]{0.45\linewidth}
\centering 
     \includegraphics[height=4.0cm]{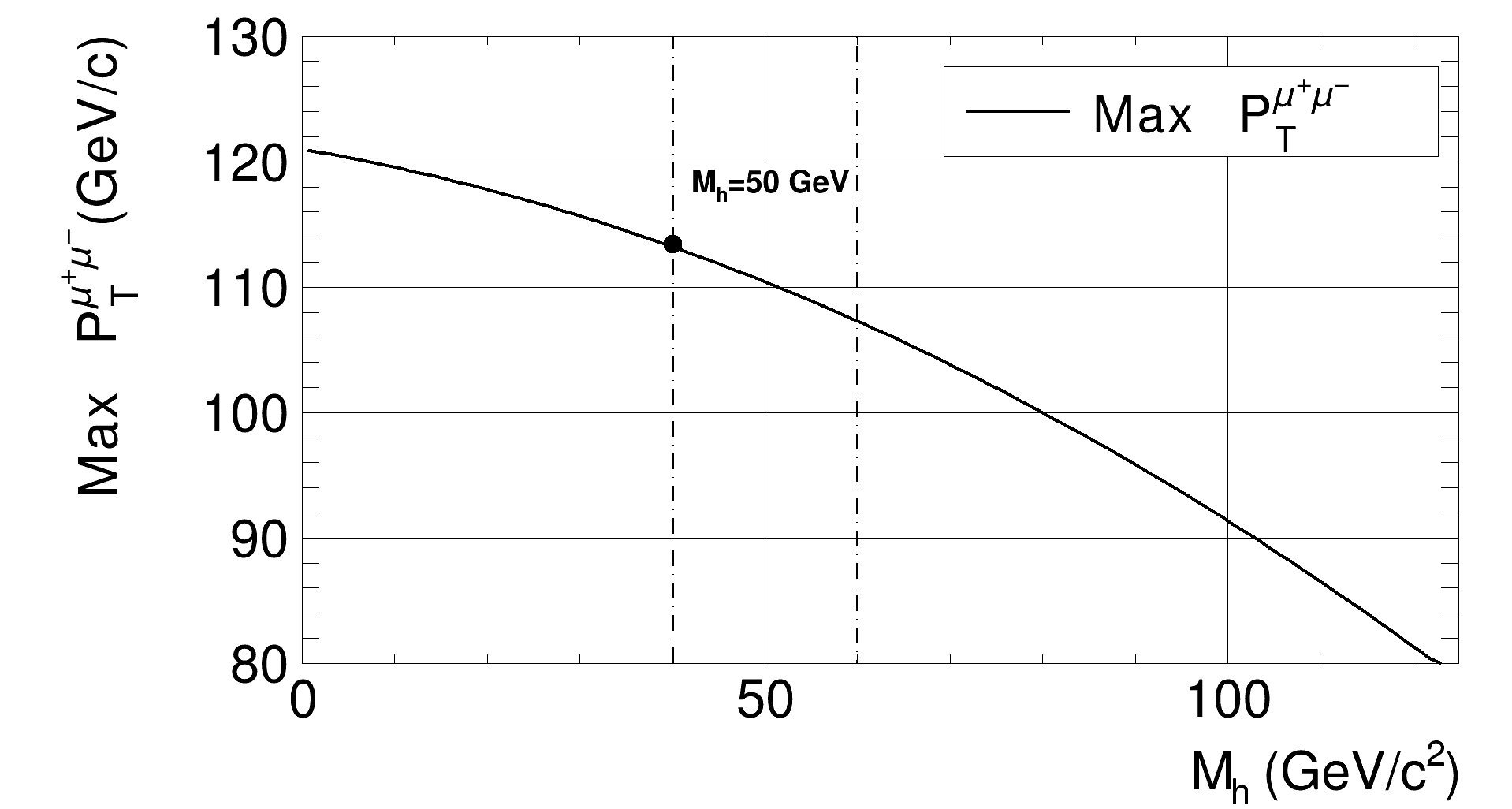}		
     
     (a)
\end{minipage}
\hspace{1cm}
  \begin{minipage}[t]{0.45\linewidth}
      \centering
     \includegraphics[height=4.5cm]{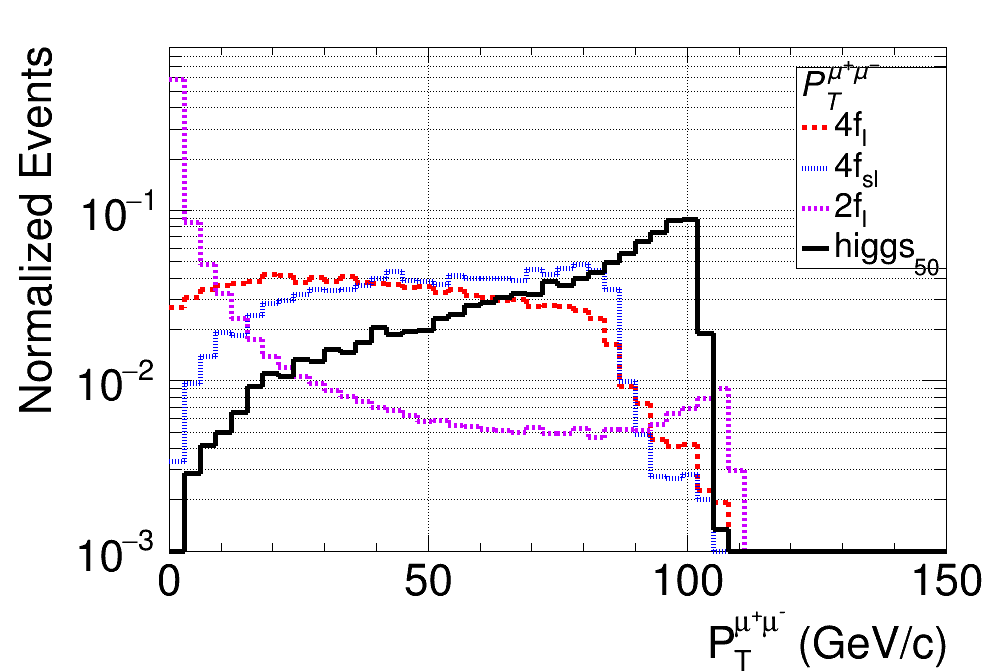}		
     
     (b)
\end{minipage} 
     \caption{(a) Maximum value for transverse momentum of muon pair system as a function of higgs mass; (b) transverse momentum distribution of muon pair system for $\mh=50\gev$.}\label{f_zpt}	
\end{figure} 

\begin{table}[h]
   \begin{center}
 \begin{small}  
 \begin{tabular}{|c|c|c|c|c|c|} 
 \hline
 $\mh$ (GeV) & 30  & 50  & 70  & 90  & 115  \\
 \hline
 Max $\ptmu$ cut (GeV)  & 120 & 115 & 110 & 100 & 90 \\
 \hline
 \end{tabular}
  \end{small}
   \caption{Transverse momentum uplimits of muon pair system for different higgs masses.}\label{t_cut_zpt}
 \end{center}
 \end{table}

  \item  The $2f_l$ processes contain a large number of ISR photons. If photons escape through the beam pipe, they contribute to the missing momentum in longitudinal direction. Thus, the cut for $\cosmis<$ 0.98 can reject many $2f_l$ background events.
However, when the higgs mass is small, more ISR photons will escape from the detector. Thus, the $|\cosmis|<0.98$ cut is applied only for $\mh>50\gev$.

 \item The signatures of the $ZZ$ background with one $Z$ boson decay to a muon pair are harder to distinguish from the signal. Therefore, a multi-variate analysis (MVA) based on the Gradient Boosted Decision Tree (BDTG) method \cite{QUINLAN1987221}, which is included in TMVA package \cite{BRUN199781} in ROOT \cite{Hocker:2007ht}, is used for further background rejection. The input variables  for the BDTG are $\text{cos}\theta_{\mu^{+}\mu^{-}}$, $\text{cos}\theta_{\mu-\mu}$, $\text{cos}\theta_{\mathrm{track+}}$,
$\text{cos}\theta_{\mathrm{track-}}$ and $\Mmu$. Here, $\theta_{\mu^{+}\mu^{-}}$ is the polar angle of the $Z$ boson, $\theta_{\mu-\mu}$ is the angle between the muons, and $\theta_{\mathrm{track+/-}}$ is the polar angle of $\mu^{+}/\mu^{-}$ track. Figure \ref{f_BDTG_input}  shows distribution of these variables for $\mh=50\gev$. 

The BDTG is trained by using half of  simulated signal and background events. Then the BDTG response is calculated with the other half of the events. In most $\mh$ cases, there are no obvious changes in final sensitivity for BDTG cuts between (-0.3, 0.3).  The BDTG cut is chosen to be larger than 0 for all higgs masses. Figure \ref{f_BDTG} shows as an example the BDTG response for $\mh=50\gev$. 

\begin{figure}[h]
\begin{minipage}[t]{0.45\linewidth}
\centering
\includegraphics[height=4.5cm]{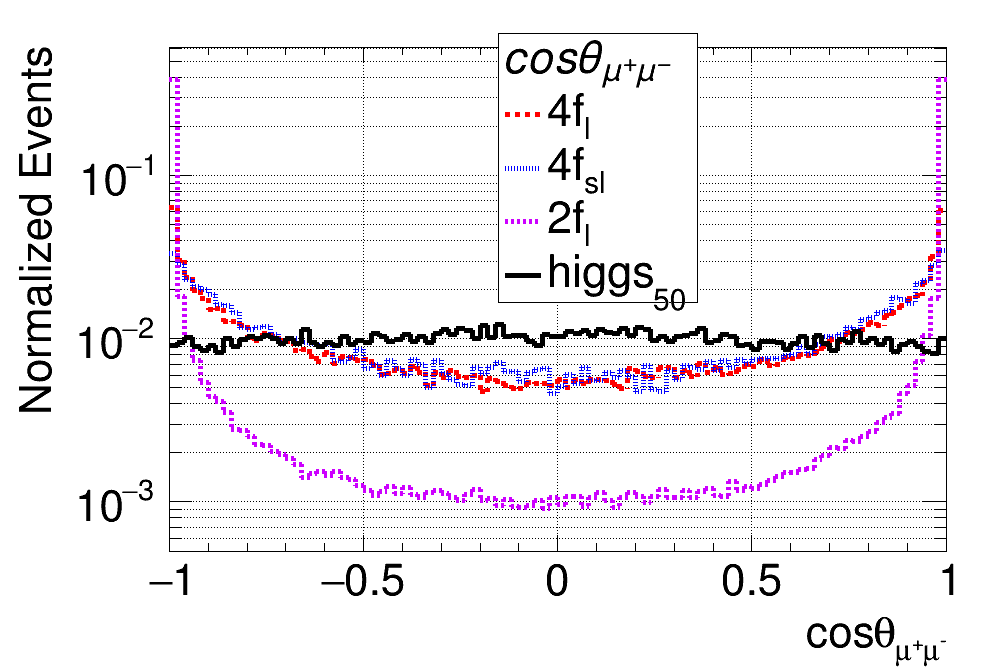}\\
\end{minipage}
\begin{minipage}[t]{0.45\linewidth}
\centering
 \includegraphics[height=4.5cm]{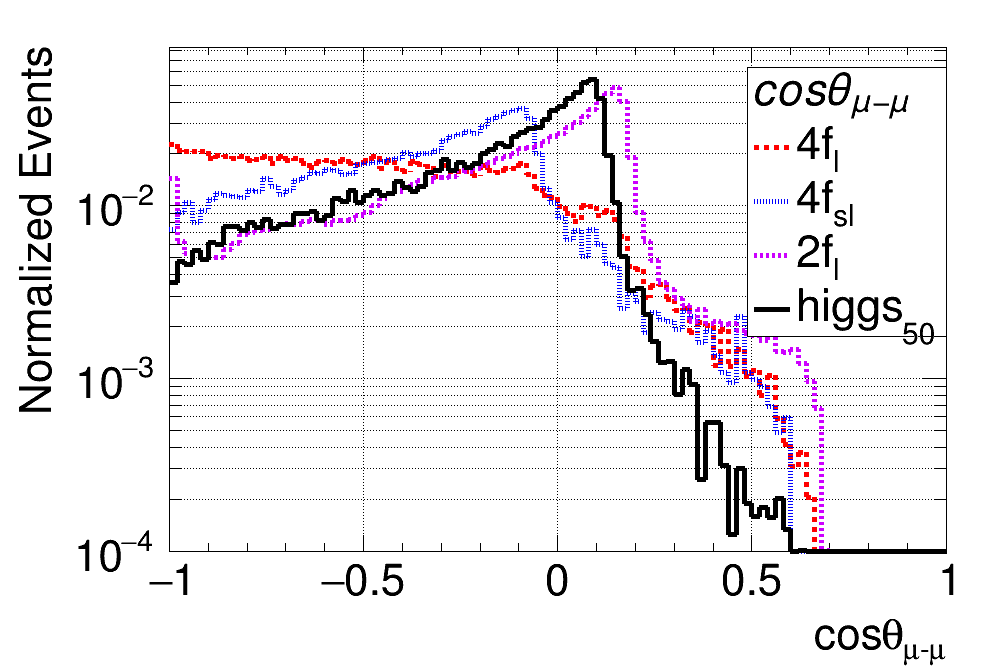}	
\end{minipage}

  \begin{minipage}[t]{0.45\linewidth}
\centering
\includegraphics[height=4.5cm]{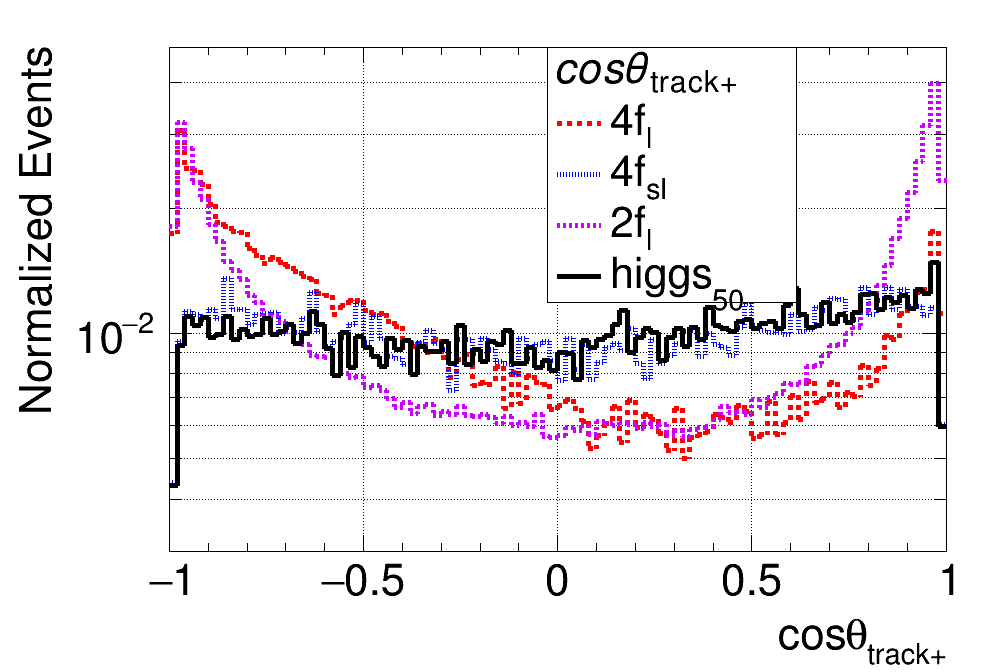}\\
\end{minipage}
  \begin{minipage}[t]{0.45\linewidth}
\centering
 \includegraphics[height=4.5cm]{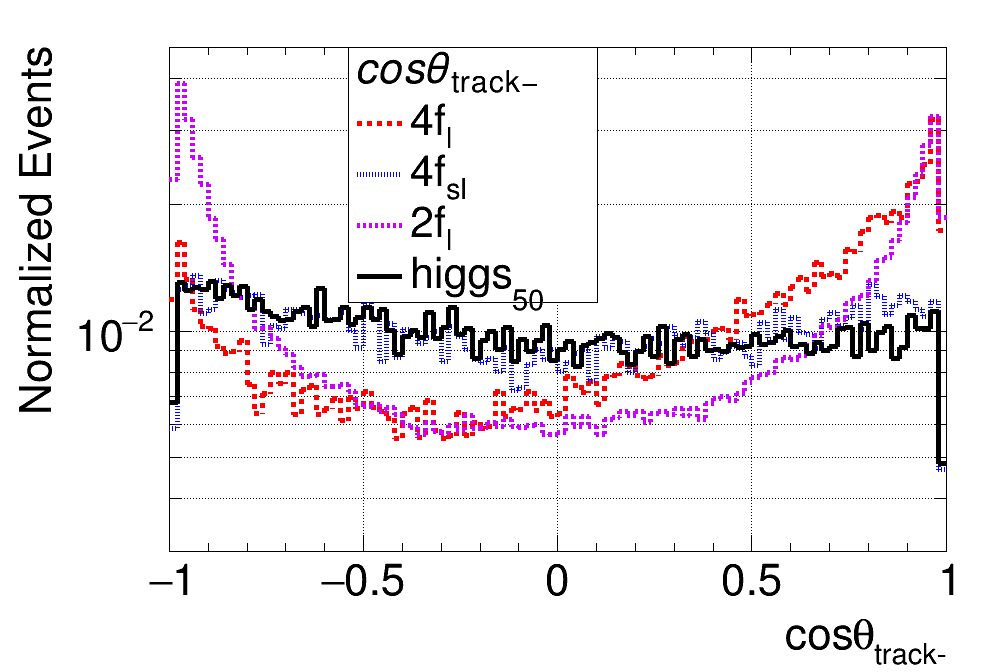}
\end{minipage}
     \caption{Example distributions of the variables  $\text{cos}\theta_{\mu^{+}\mu^{-}}$, $\text{cos}\theta_{\mu-\mu}$,  $\text{cos}\theta_{\mathrm{track+}}$ and  $\text{cos}\theta_{\mathrm{track-}}$ for $\mh=50\gev$, which is used for the BDTG training input variables.}\label{f_BDTG_input}
\end{figure}

\begin{figure}[h]
\centering
  \includegraphics[height=6cm]{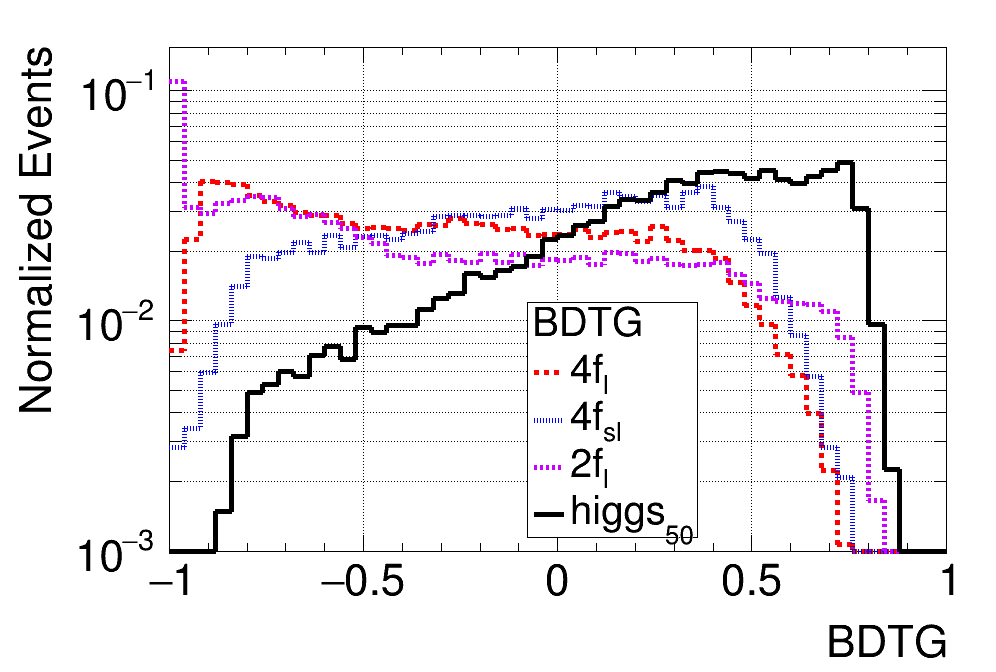}	
  \caption{The distribution of the BDTG response for signal and background for $\mh=50\gev$.}\label{f_BDTG}
\end{figure}

\item The unique characteristic of the signal is the peak of the recoil mass close to the higgs mass. This motivates the cut $[(\mh-20),160]\gev$. In Figure \ref{f_recoil_eg}, taking $\mh=50\gev$ for example, it shows the recoil mass distribution for signal and backgrounds.  
\begin{figure}[h]
\centering
  \includegraphics[height=6cm]{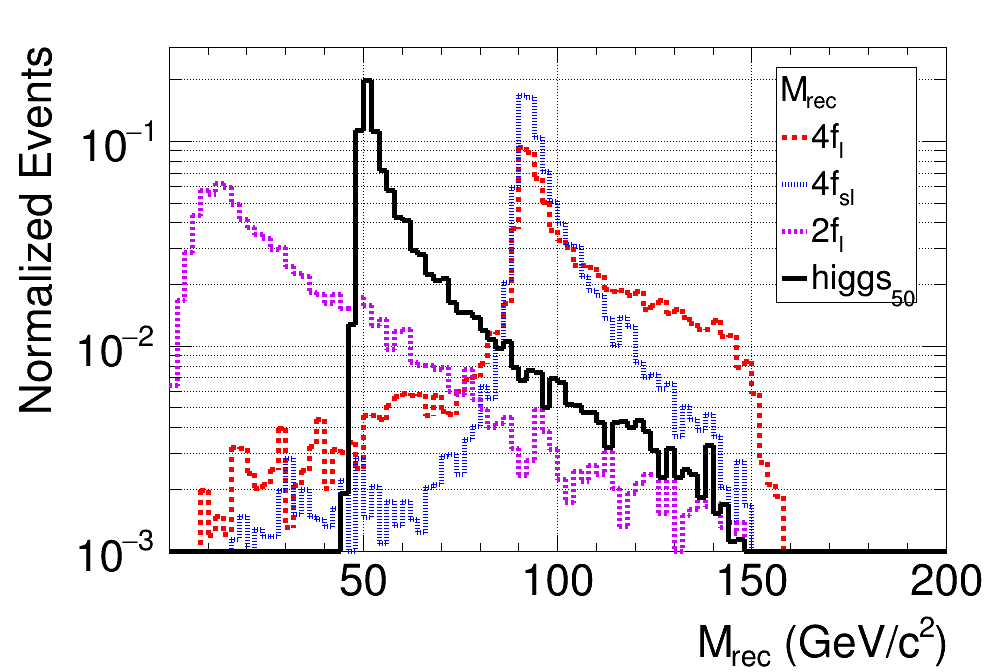}	

     \caption{The recoil mass distribution of the signal and background for $\mh=50\gev$.}\label{f_recoil_eg}
\end{figure}

\end{itemize}

According to the recoil mass distribution, the main backgrounds are distinguishing for different higgs boson masses. The signal-to-background ratio is clearer in  $125\gev>\mh>\mz$ and $\mz>\mh>50\gev$ regions, because no background processes have the same feature as the signal. While in the $\mh\sim \mz$ region, the $ZZ$ process with one $Z$ boson decaying to a muon pair is an irreducible background; and in $\mh<50\gev$  region, the $2f_{l}$ background $e^{+}e^{-} \to Z \to \mu^{+}\mu^{-}$ with energetic ISR photon is overwhelming the signal. 

%\begin{table}[h]
%\begin{center}
% \begin{tabular}{|c|c|} \hline
% mass region & main backgrounds \\
% \hline
%$125>M_{h}>M_{Z}$  & $4f_{zz}^{sl},\quad 4f_{zzww}$ \\
% \hline
%$M_{h}\sim M_{Z}$  & $4f_{zz}^{l}, \quad 4f_{zz}^{sl}, \quad 4f_{zzww}$ \\
% \hline
%$M_{Z}>M_{h}>40$  & $2f_{l}, \quad 4f_{zz}, \quad 4f_{zzww}$ \\
% \hline
%$40>M_{h}$  & $2f_{l}$\\
% \hline
% \end{tabular}
% \end{center} 
%     \caption{The main backgrounds for different higgs mass regions.}\label{t_signal_regioin}
% \end{table}
%  

\section{Results}\label{s_result}

In Table \ref{t_data}, there are the number of remaining signal and background as well as the signal efficiency and significance after all cuts for different higgs masses. The significance is defined as $\frac{S}{\sqrt{S+B}}, \quad \text{and}~S = \kappa_{hZZ}^2 \times\sigma_{h\mu\mu}^{\mh}  \times \int L dt, ~\text{where}~\kappa_{hZZ}=\,1$. The cross section $\sigma_{h\mu\mu}^{\mh}$  increases when the scalar mass becomes small due to kinematics, which leads to a higher significance.  
  \begin{table}[h]
 \begin{center}
 \begin{tabular}{|c|c|c|c|c|c|c|c|} \hline
$\mh$(\gev~)& light higgs  & $4f_{l}$  & $4f_{sl}$  & $2f_{l}$  & $\text{total~bk}$  & \thead{cut \\ efficiency} & significance \\
 \hline
$115$  & 17420 & 61034 & 53869 & 13878 & 128781 & 0.67 & 45.56\\
\hline
$90$   & 22198 & 63211 & 74563 & 18514 & 156288 & 0.59 & 52.54\\
\hline
$70$   & 26841 & 51672 & 60358 & 37167 & 149196 & 0.57 & 63.97\\
\hline
$50$   & 30494 & 46128 & 54373 & 80074 & 180575 & 0.54 & 66.37\\
\hline
$30$  & 33844 & 51207 & 55743 & 213184 & 320134 & 0.49 & 56.88\\
 \hline
 \end{tabular}
 \end{center}
      \caption{The number of events left after all kinematic cuts for $P(e^{-},e^{+})$ = $(-80\%,\,+30\%)$, $\int L dt =2000 \fb^{-1} $ and $\sqrt{s} = 250\gev$.  Also given are the efficiency and signal significance (defined as $\text{significance} = \frac{S}{\sqrt{S+B}}, \quad \text{and}~S = \kappa_{hZZ}^2 \times\sigma_{h\mu\mu}^{\mh}  \times \int L dt,~\text{where}~\kappa_{hZZ}=1$).}\label{t_data}
 \end{table}

A likelihood analysis is applied for calculating 2$\sigma$ expected exclusion limits on the coupling $\kappa_{hZZ}$ with a bin-by-bin comparison between the  signal and background recoil mass histograms. 
Two hypotheses are provided, the background-only hypothesis which assumes no new higgs in the investigated mass range, while the signal-plus-background hypothesis which assumes the new higgs is produced in the mass range.
Then a global test-statistic $Q(\mh)=\mathcal{L}_{s+b}(s(\mh))/\mathcal{L}_{b}(0)$ is constructed to discriminate signal and background, where the $\mathcal{L}_{s+b}~(\mathcal{L}_{b})$ is the likelihood function for the signal plus background (only background) hypothesis.
The distribution of $Q(\mh)$ is normalized so that it becomes a probability density function, which is integrated to provide the confidence levels $CL_b(\mh)$ and $CL_{s+b}(\mh)$.
The ratio $CL_s(\mh)=CL_{s+b}(\mh)/CL_b(\mh)$ is used as the final confidence level.

\begin{figure}[h]
\begin{center}
     \includegraphics[height=6cm]{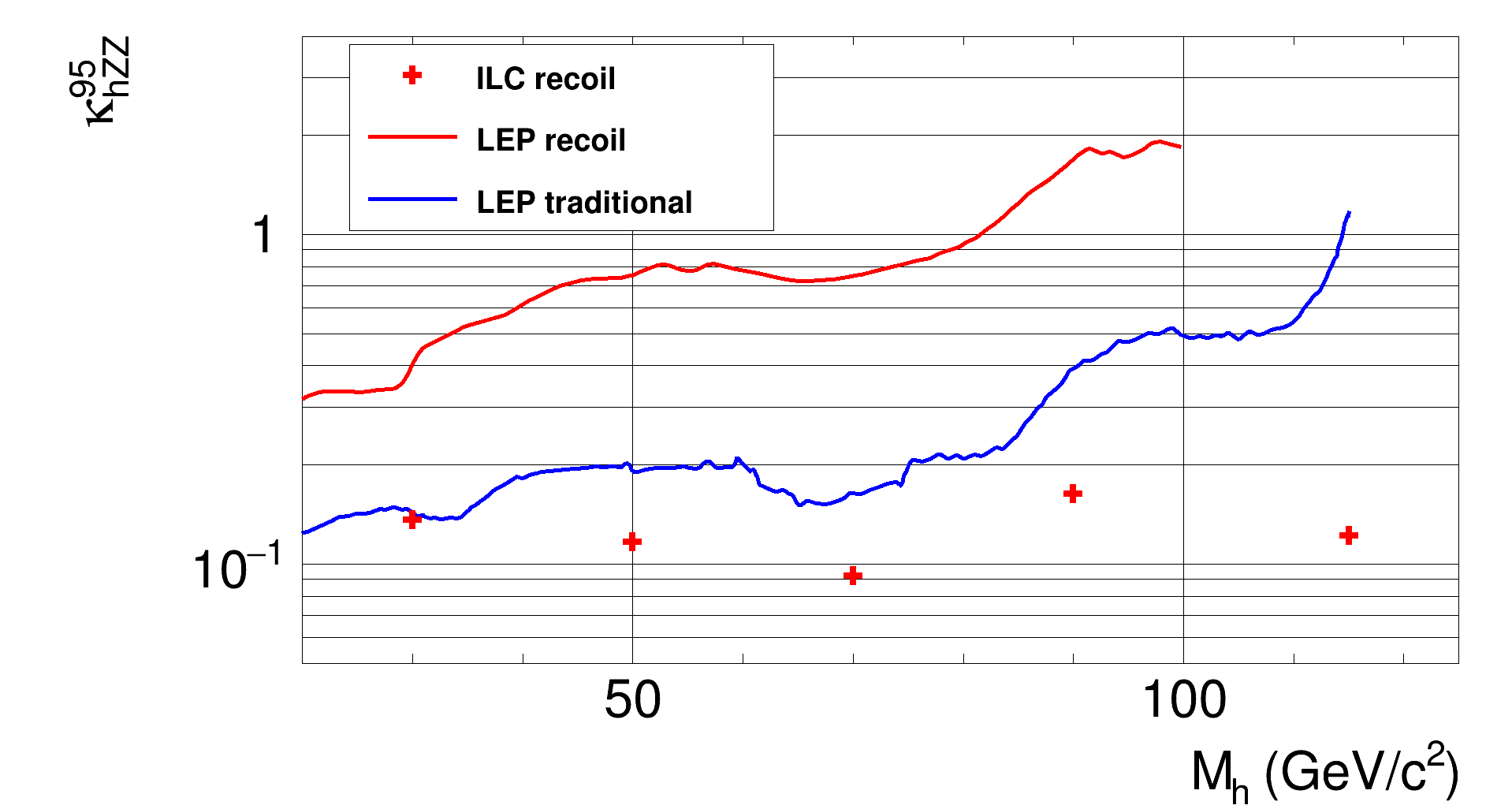}		
\end{center}
     \caption{The 2$\sigma$ exclusion limits for the coupling $\kappa_{hZZ}$ for different higgs masses. The black points are the results at ILC with  polarization $P(e^{-},e^{+})$ = $(-80\%,\,+30\%)$, $\int L dt =2000 \fb^{-1} $ and $\sqrt{s} = 250\gev$. The green line is the results with recoil technique at LEP, while the red line use traditional methods.}\label{f_exclusive}
\end{figure}

Finally, the 95\% confidence level upper bounds on $\kappa_{hZZ}$ ($\kappa_{hZZ}^{95}$) is calculated for five higgs mass benchmark points, which is shown in Figure \ref{f_exclusive} with the black points. 
Comparing the $\kappa_{hZZ}^{95}$ in $\mh\sim \mz$ and $40\gev>\mh$ regions, the coupling magnitudes reach the valley bottom in $125\gev>\mh>\mz$ and $\mz>\mh>40\gev$ regions.

Figure \ref{f_exclusive} also shows the results at LEP for comparison. 
The red line is $\kappa_{hZZ}^{95}$ measured at LEP by combining the data of the four LEP collaborations, ALEPH, DELPHI, L3 and OPAL \cite{Barate:2003sz}. In each of the four LEP experiments, the data analysis is done with traditional methods, $i.e.$ identifying the decay modes of the higgs bosons in the discrimination between signal and background. 
The  green line presents $\kappa_{hZZ}^{95}$ obtained with the recoil technique by OPAL Collaboration \cite{Abbiendi:2002qp} at LEP. In this search, the scalar masses have been measured down to the lowest generated signal mass of 1 keV, which is the only model-independent higgs search at LEP due to limited luminosity. The $\kappa_{hZZ}^{95}$ is independent of the decay modes of the higgs boson, which is $\kappa_{hZZ}^{95} < 1$ when $\mh < 81\gev$.  When extrapolating these results to  the ILC with $\sqrt{s}=250\gev$, $P(e^{-},e^{+})~=~(-80\%, \,+30\%)$ and $\int L dt=2000\fb^{-1}$, the $\kappa_{hZZ}^{95}$ is estimated to be $[0.055- 0.071]$ \cite{lcws_peter}. These extrapolated exclusion limits are about a half of the values given in this analysis, especially in the low mass region. The main reason is that the OPAL analysis has been split into visible and invisible decay modes of the higgs boson, and assume that the signal events rejected in the visible decay mode analysis (by photon rejection) can be recovered in the invisible decay study (including mono-photon production in signal decay channel)\cite{Abbiendi:2002qp}. This may slightly sacrifice model independence. The current ILC results are completely model-independent. If the slight model dependence can be accommodated, the ILC can   also reach the extrapolated exclusion limits. Another reason is the center-of-mass energy of ILC is larger than LEP. Thus, when higgs mass is small, the recoil mass distribution is duller than that in LEP, which makes it harder to reject the backgrounds by recoil mass cut. 
\section{Conclusions}\label{s_conclusion}

Many BSM models  favor light scalars. By applying the recoil technique, the potential of the ILC to search for scalars has been investigated  at $\sqrt{s}=250\gev$, with the full simulation of the ILD detector concept. 
We have optimized the methods of signal selection and background rejection
to be independent of the scalar decay modes.
Preliminary 2$\sigma$ expected exclusion limits for scale factor $\kappa_{hZZ}^{95}$ of the $hZZ$ coupling are shown for five scalar mass benchmark points.
The analysis shows two irreducible backgrounds affect the final results, two fermion backgrounds will be dominant in the low higgs mass region, while the $ZZ$ process with one $Z$ boson decaying to a muon pair is irreducible when $\mh$ is close to $Z$ boson mass. 
In future, we plan to split the decay mode of higgs boson into visible and invisible to improve the background rejection and extend our analysis to more mass points and check the model independency for each higgs decay modes. 

\Acknowledgements
We would like to thank the LCC generator working group and the ILD software working group for
providing the simulation and reconstruction tools and producing the Monte Carlo samples used in this
study. This work has benefited from computing services provided by the ILC Virtual Organization,
supported by the national resource providers of the EGI Federation and the Open Science GRID. 
We are grateful for the support from Collaborative Research Center SFB676 of the Deutsche Forschungsgemeinschaft (DFG), Particles, Strings and the Early Universe, project B1.
Y.W. is supported by the China Postdoctoral Science Foundation under Grant No. 2016M601134, and an International Postdoctoral Exchange Fellowship Program between the Office of the National Administrative Committee
of Postdoctoral Researchers of China (ONACPR) and DESY.

\printbibliography

\end{document}